\documentclass[aps,prl,reprint,superscriptaddress,showpacs]{revtex4-1}
\usepackage[latin9]{inputenc}
\usepackage{babel}
\usepackage{amsmath}
\usepackage{amssymb}
\usepackage{wrapfig}
\usepackage{graphicx}
\usepackage[unicode=true,
 bookmarks=false,
 breaklinks=false,pdfborder={0 0 1},colorlinks=true,linkcolor=red,citecolor=blue]
 {hyperref}
\makeatletter
\@ifundefined{textcolor}{}
{%
 \definecolor{BLACK}{gray}{0}
 \definecolor{WHITE}{gray}{1}
 \definecolor{RED}{rgb}{1,0,0}
 \definecolor{GREEN}{rgb}{0,1,0}
 \definecolor{BLUE}{rgb}{0,0,1}
 \definecolor{CYAN}{cmyk}{1,0,0,0}
 \definecolor{MAGENTA}{cmyk}{0,1,0,0}
 \definecolor{YELLOW}{cmyk}{0,0,1,0}
}

\usepackage{amsthm}
\usepackage{amsfonts}
\usepackage{bbm}
\usepackage{epsfig}
\usepackage{times}
\usepackage{babel}
\usepackage{color}
\usepackage{framed}
\usepackage{changes}
\usepackage{float}
\usepackage{array}
\usepackage{geometry}
\usepackage{subfig}
\renewcommand{\raggedright}{\leftskip=0pt \rightskip=0pt plus 0cm}

\makeatother

\begin{document}

\title{The length of a compact extra dimension from black hole shadow}

\author{Zi-Yu Tang}
\email{tangziyu@ucas.ac.cn}
\affiliation{School of Fundamental Physics and Mathematical Sciences, Hangzhou Institute for Advanced Study, UCAS, Hangzhou 310024, China}
\affiliation{School of Physical Sciences, University of Chinese Academy of Sciences, Beijing 100049, China}

\author{Xiao-Mei Kuang}
\email{xmeikuang@yzu.edu.cn}
\affiliation{Center for Gravitation and Cosmology, College of Physical Science and Technology, Yangzhou University, Yangzhou 225009, China}

\author{Bin Wang}
\affiliation{School of Aeronautics and Astronautics, Shanghai Jiao Tong University, Shanghai 200240, China}
\affiliation{Center for Gravitation and Cosmology, College of Physical Science and Technology, Yangzhou University, Yangzhou 225009, China}

\author{Wei-Liang Qian}
\affiliation{Escola de Engenharia de Lorena, Universidade de S$\tilde{a}$o Paulo, 12602-810, Lorena, SP, Brazil}

\affiliation{Faculdade de Engenharia de Guaratinguet$\acute{a}$,
Universidade Estadual Paulista, 12516-410, Guaratinguet$\acute{a}$, SP, Brazil}

\affiliation{Center for Gravitation and Cosmology, College of Physical Science and Technology, Yangzhou University, Yangzhou 225009, China}

\maketitle
The conception of extra dimension was first introduced by Nordstr\"omin, in order to unify electromagnetism and gravity \cite{Nordstrom:1914ejq}. 
Nowadays, it has become a common idea that besides our four-dimensional spacetime, the extra spatial dimension is a relevant notion in the framework of particle physics, gravity and cosmology. 
Moreover, it appears in phenomenological (bottom-up) approaches and quantum gravity models such as string theories. Among others, the five-dimensional Kaluza-Klein (KK) theory is shown to recover both the electrodynamics and general relativity (GR) in the four-dimensional spacetime, where the extra dimension is considered to be a compact circle \cite{Klein:1926fj}. 
Thereafter the domain wall theory was constructed with an infinite extra dimension and a bulk scalar field \cite{Rubakov:1983bb}, where the effective potential well along the extra dimension could localize the energy density of the scalar field on a three-dimensional hypersurface, i.e. the domain wall embedded in the five-dimensional spacetime. Nonetheless, both models have their limitations since a well-defined extra-dimensional theory should not only explain the particles in the Standard Model but also provide a resolution to the hierarchy problem (the large discrepancy between the Planck scale and the electroweak scale).  
Later, the well-known RS-I \cite{Randall:1999ee} and RS-II \cite{Randall:1999vf} models were proposed as a well-defined extra-dimensional theory, in which a warped structure was introduced to the  compact/infinite extra dimension.  
More details on these models can be found in \cite{Yu:2019jlb}.
Furthermore, higher dimensional black holes have always been attractive in the context of string theories and braneworld models. Our observable world is effectively four-dimensional, while we can treat higher dimensional black hole solutions with extra dimensions as candidates for realistic models. Gregory and Laflamme made the pioneering attempt to generalize a four-dimensional Schwarzschild black hole to a five-dimensional black string by the extension to an extra dimension with the topology $\mathbb{S}^4_{Sch}\times \mathbb{R}^1$, which can be regarded as an extra hair of black holes \cite{Gregory:1987nb}. The well-known GL instability of black strings/branes was addressed, but the instability can be evaded by the compactification of the extra dimensions \cite{Gregory:1993vy}.
To date, the essence of the extra dimension, such as the number, shape, and size, remains an open question.

Physicists used to focus the detecting extra dimensions on high-energy experiments. Before the achievement of the Gravitational Wave (GW) detection \cite{LIGOScientific:2016aoc}, which opens a new window to probe the structure of black holes, physicists made efforts to study the features of GWs in extra-dimensional theories that can distinguish the effects of extra dimensions from those in other modified gravity theories, mainly the discrete high-frequency spectrum and shortcuts, see a recent review \cite{Yu:2019jlb}. However, the discrete high-frequency spectrum (about $\ge 300$ GHz) is far beyond the scope of GW detectors at present, and not all GWs in extra-dimensional theories can take shortcuts, so the information of extra dimensions from current GW detection is still difficult. Besides GW detection, the other recent breakthrough of detecting strong field regime of black hole is from the Event Horizon Telescope (ETH) collaborations, which published the first black hole photo of M87* in 2019  \cite{EventHorizonTelescope:2019dse} and the second one for SgrA* in 2022 \cite{EventHorizonTelescope:2022xnr}.
In this work, we propose to use the black hole shadow observations from EHT to explore the extra dimensions, and find a suitable way to constrain its size.

The deflection of light near a massive object has been verified since 1919, as a prediction of GR. While in the spacetime near a black hole, the gravity will be so strong that the photons can orbit the black hole and form photon region. Usually such photon orbits are unstable, and slight deviation will make the photons drop into the black hole or run away to infinity. Therefore a black hole looks like a dark disk surrounded with a shine doughnut, as shown in the black hole photos published by the EHT Collaboration \cite{EventHorizonTelescope:2019dse, EventHorizonTelescope:2022xnr}. The observation of the black hole shadow provides direct information of the geometry near the event horizon of the black hole, and develops as a new research focus in recent years \cite{Perlick:2021aok}.

\begin{figure*}
\begin{centering}
{\includegraphics[width=.32 \textwidth]{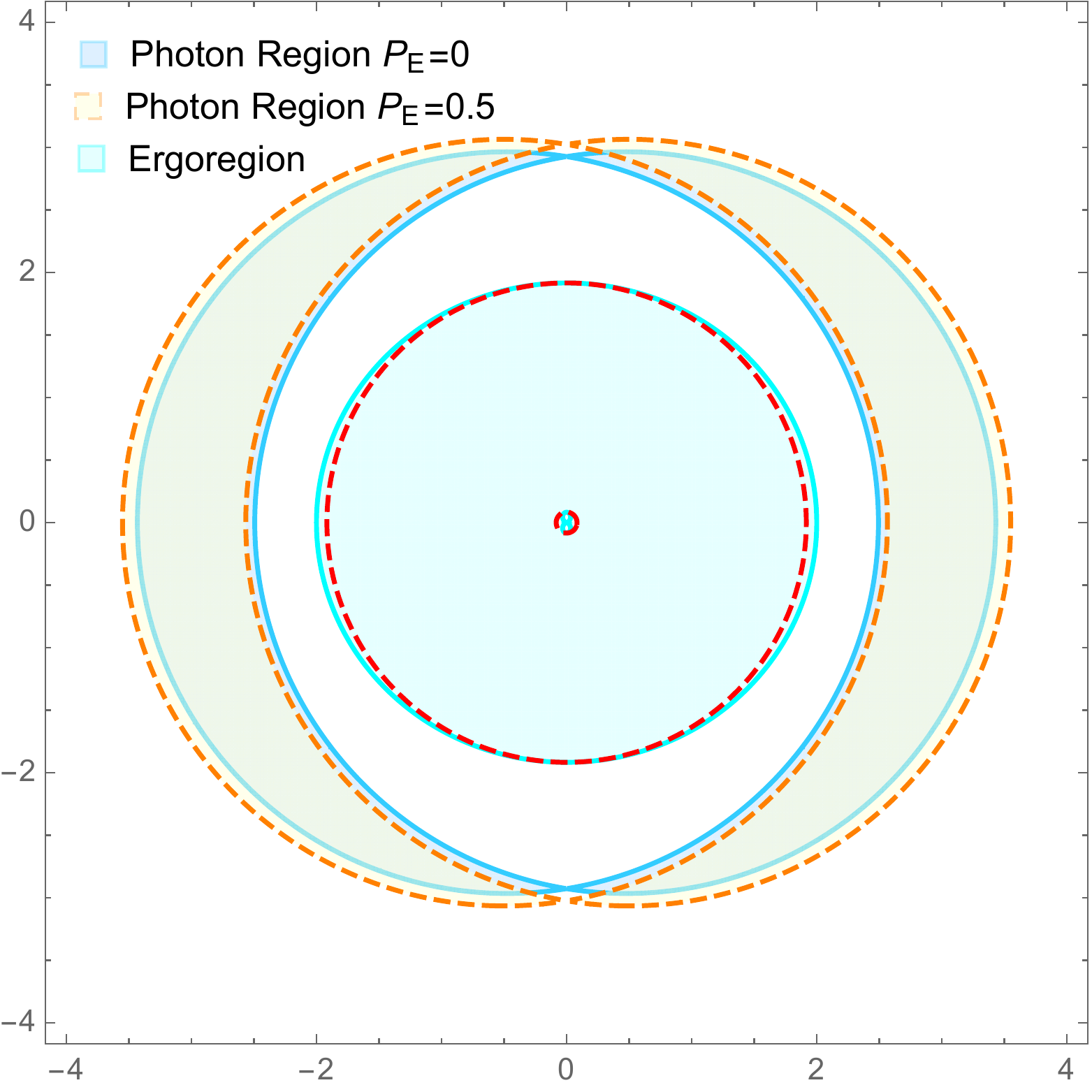}}\hspace{1cm}
{\includegraphics[width=.34\textwidth]{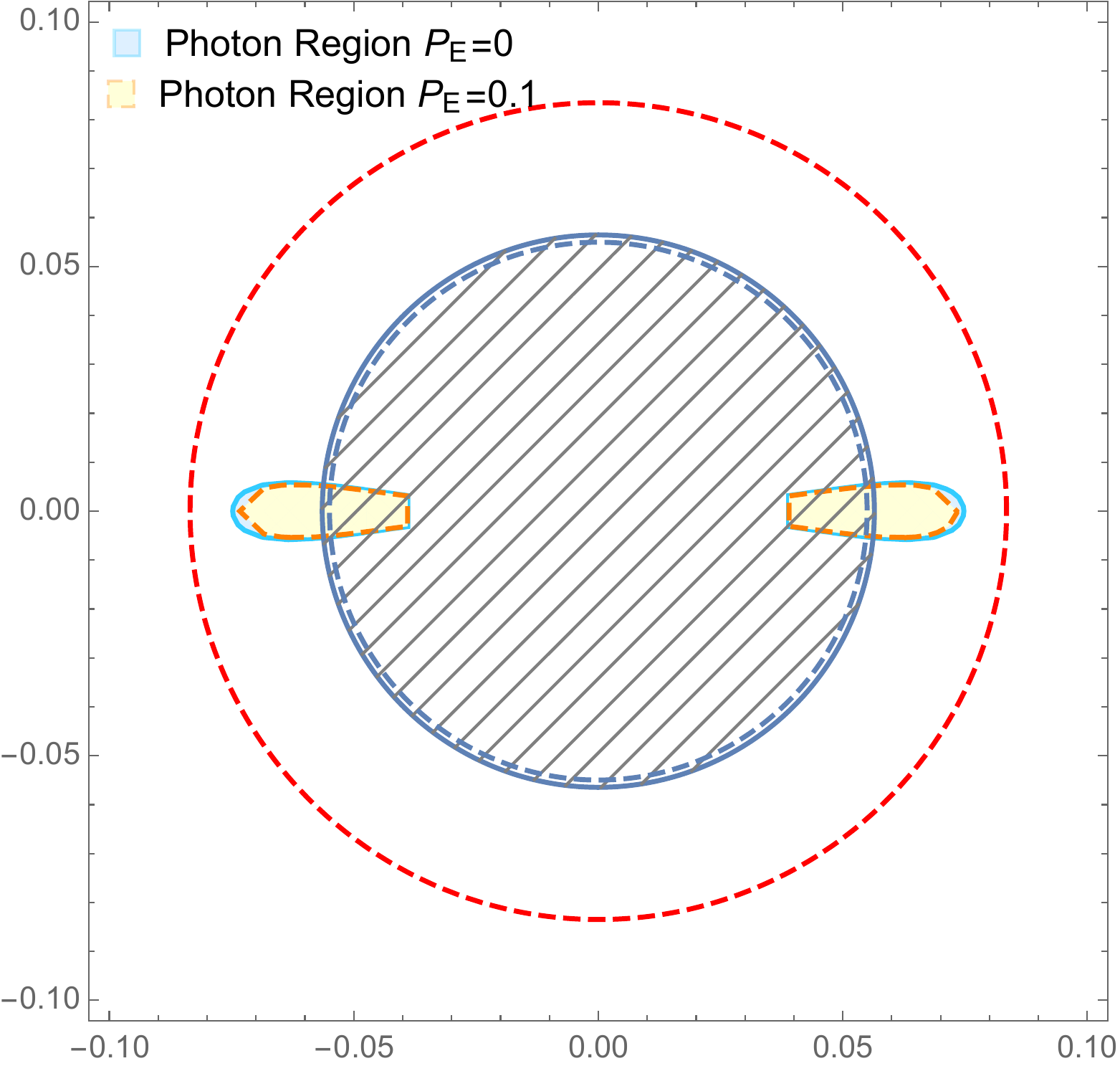}}\hspace{-2.5cm}
\newline \\
{ \includegraphics[width=0.35 \textwidth]{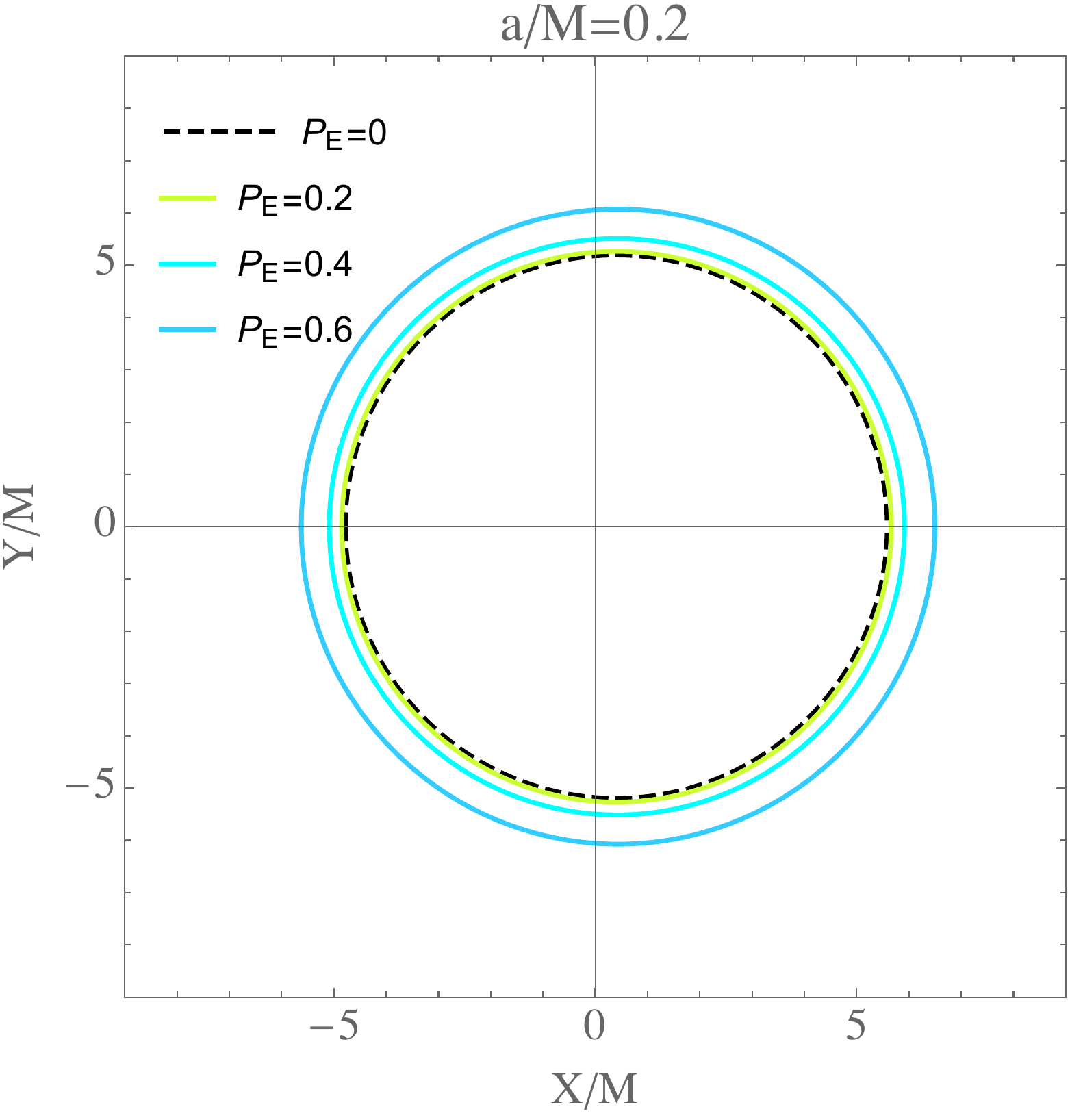}}\hspace{1cm}
{ \includegraphics[width=0.35 \textwidth]{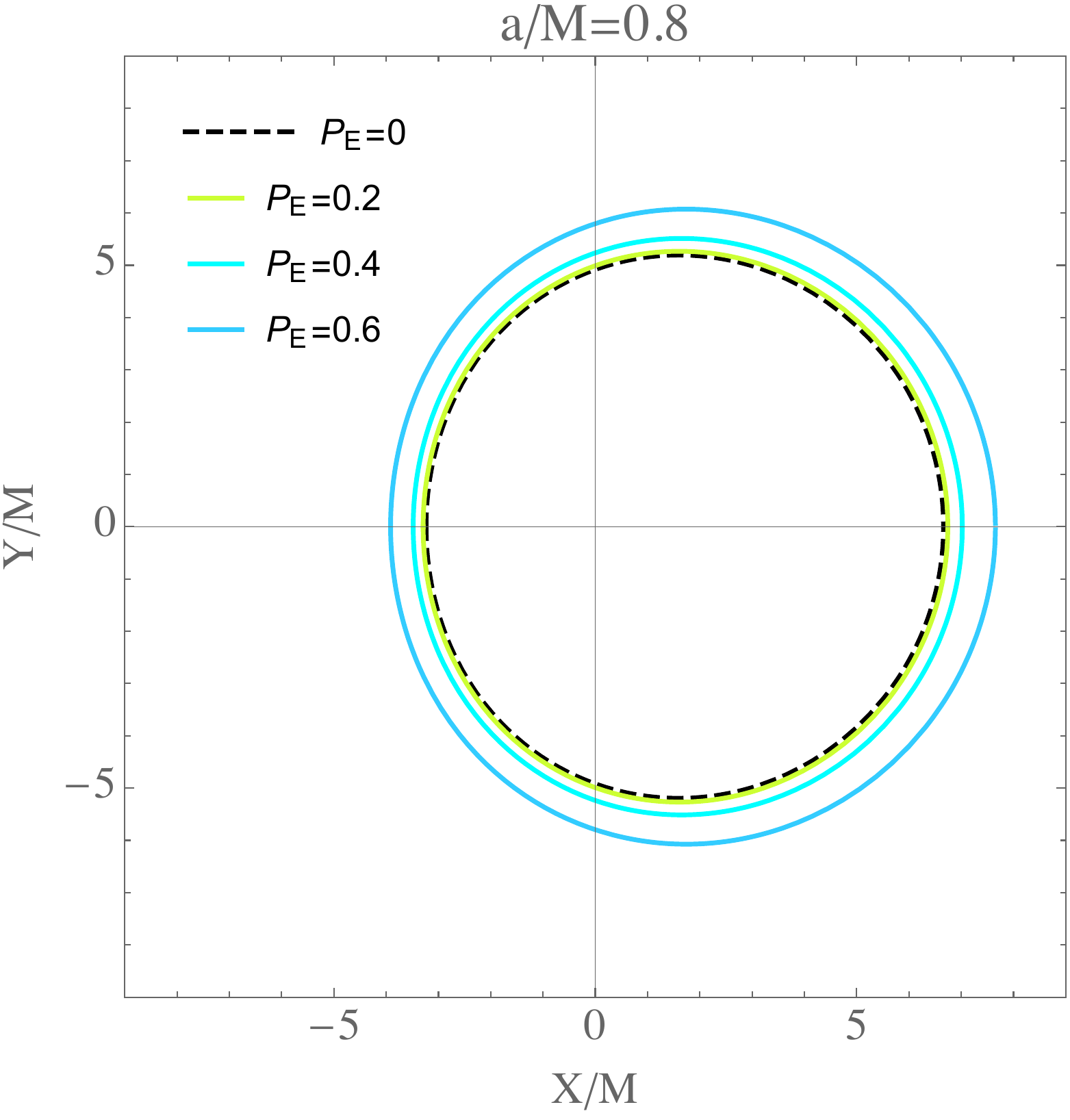}}\hspace{0.5cm}
\par \end{centering}
\captionsetup{justification=raggedright}
\caption{\footnotesize{(Color online) The photon regions outside the event horizon (upper left panel) and inside the Cauchy horizon (upper right panel) are plotted with the mass density $M=1$ and spin $a=\frac{2}{5}a_{\rm max}$ of the black string. The parameter $P_{\rm E}$ does not affect the positions of horizons (red dashed circles) and ergoregion (cyan region). Besides, the stable regions (upper right panel) for the spherical photon orbits are filled with oblique lines, where the solid boundary is for $P_{\rm z}/E_{\rm 0}=0$ and the dashed boundary is for $P_{\rm z}/E_{\rm 0}=0.1$. In the bottom, the shadow boundary of five dimensional rotating black string for a spatially infinite observer at equatorial plane is depicted with $a/M=0.2$ (left) and $a/M=0.8$ (right), respectively. The black dashed curve represents the Kerr black hole  case.}}
    \label{Fig:ShadowXY}
\end{figure*}

In general, astrophysical black holes are rotating and uncharged, which can be described by Kerr metric in four dimensions. When an extra spatial dimension $z$ is introduced in the simplest uniform way, the constructed spacetime is still a solution of the vacuum Einstein equations of GR in five dimensions \cite{Grunau:2013oca}
\begin{eqnarray}
    {\rm d}s^2=&-\frac{1}{\Sigma}\left(\Delta-a^2\sin^2{\vartheta}\right){\rm d}t^2+
    \frac{\Sigma}{\Delta}{\rm d}r^2-\frac{4a M r}{\Sigma}\sin^2{\vartheta}{\rm d}t{\rm d}\varphi\nonumber\\&+\Sigma {\rm d}\vartheta^2+\frac{1}{\Sigma}\left(\rho^4-\Delta a^2\sin^2{\vartheta}\right)\sin^2{\vartheta}{\rm d}\varphi^2+{\rm d}z^2~, 
\end{eqnarray}
where $\Sigma=r^2+a^2\cos^2{\vartheta}$, $\Delta=r^2-2Mr+a^2$ and  $\rho^2=r^2+a^2$.
Such solution describes a rotating uniform black string with an  {infinite extra dimension} $\mathbb{M}^4_{Kerr}\times \mathbb{R}^1$ or with a  {compact extra dimension} $\mathbb{M}^4_{Kerr}\times \mathbb{S}^1$, in which the length  of the black string is treated  to be the same with the length of the extra dimension. Here we start from the simplest rotating five dimensional  black strings in GR which could be regarded as a proper approximation for the realistic situation since the strict upper bounds on KK gravity with current solar system experiments and observations indicate that the extra dimension is closely flat \cite{Deng:2015sua}. Throughout the report we use the units $c=G=\hbar=1$, unless the units are specifically mentioned.

\begin{figure*}
\begin{centering}
    \includegraphics[width=.4 \textwidth]{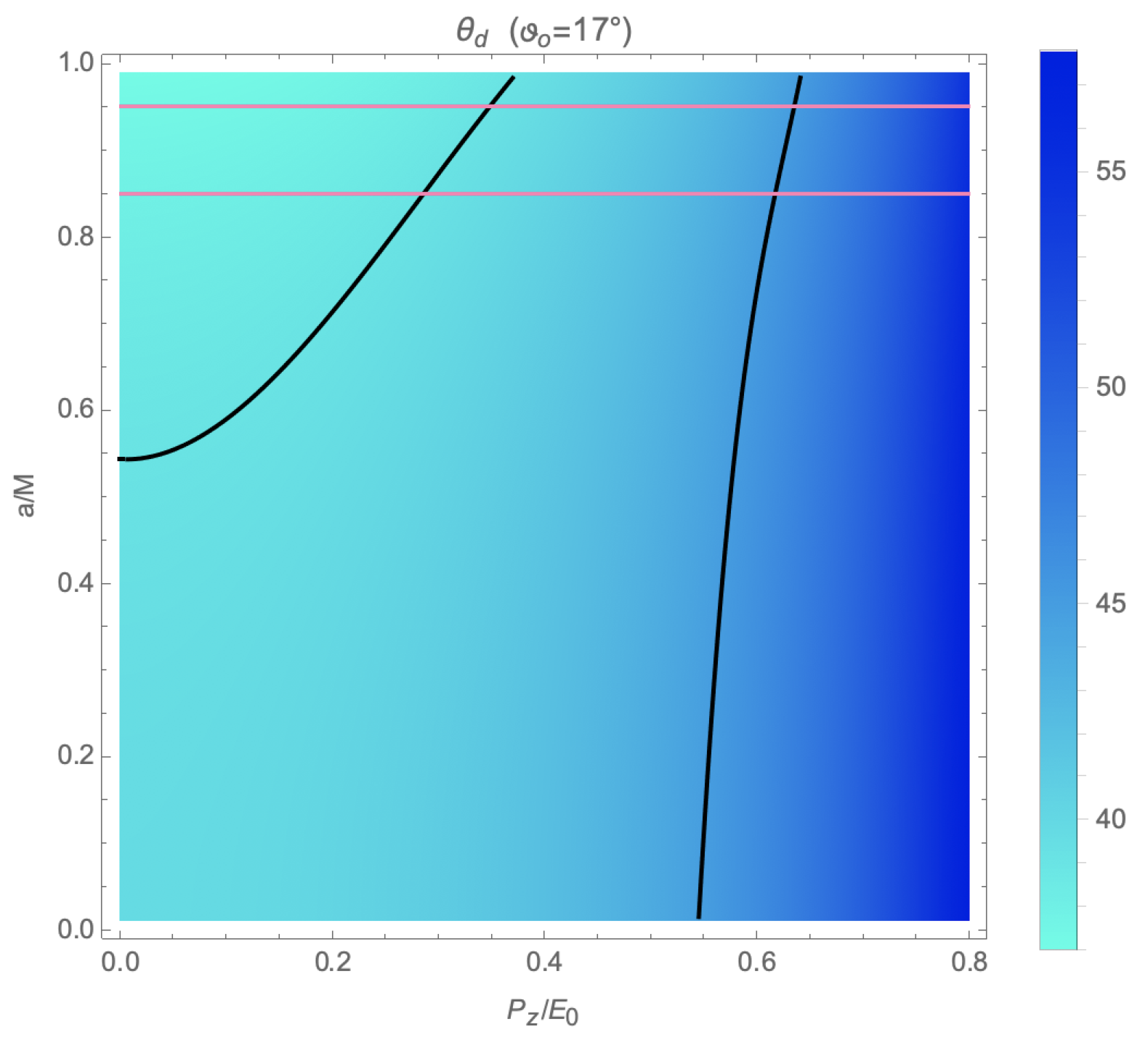}\hspace{1cm}
    \includegraphics[width=.4 \textwidth]{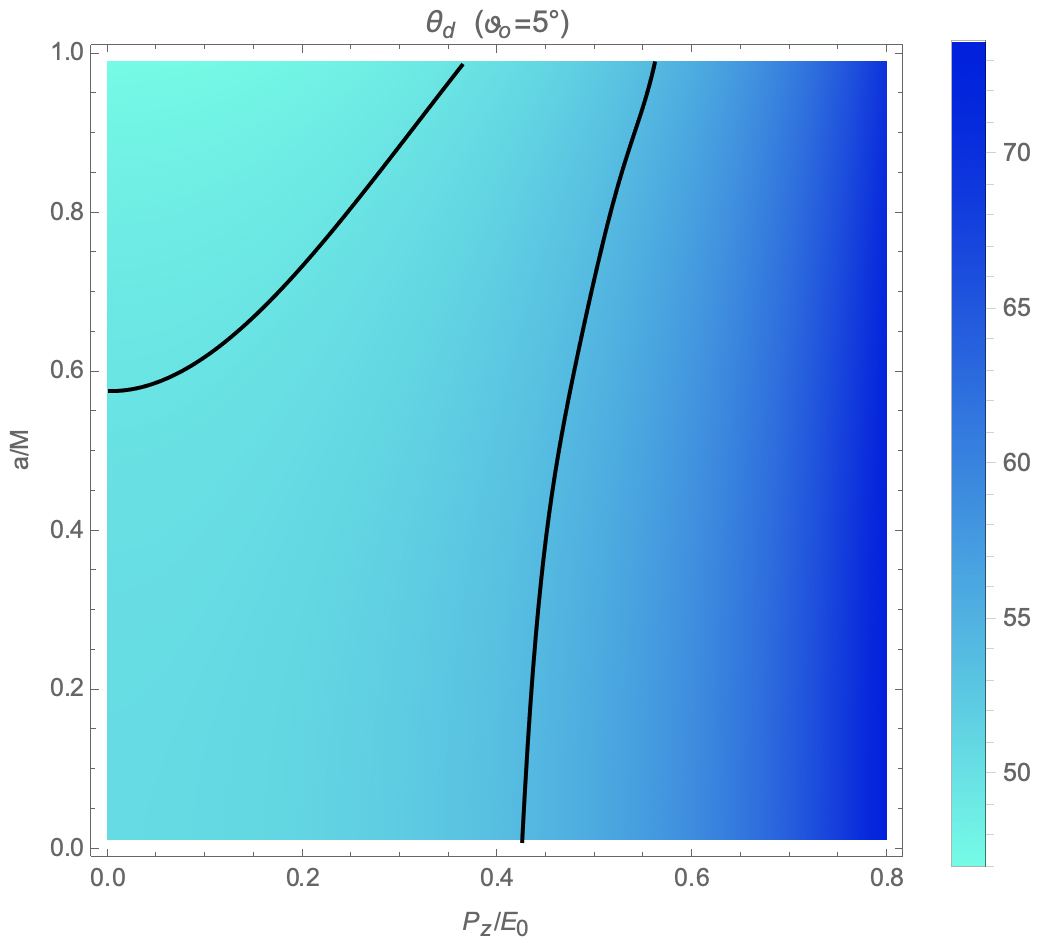}
\par\end{centering}
\captionsetup{justification=raggedright}
\caption{\footnotesize{(Color online) The shadow angular diameters of the five dimensional rotating black string shadow are plotted for supermassive M87* (left panel) with $M=(6.5\pm 0.7 )\times 10^9 M_\odot$, $d=16.8\pm 0.8 Mpc$ and $\vartheta_{\rm o}=17^\circ$ from the orientation of the jets and for supermassive SgrA* (right panel) with $M=4.0\times 10^6 M_\odot$, $d=8121~pc$ and the jet inclination $\vartheta_{\rm o}=5^\circ$. The black curves in the plots represent the ring diameter bounds estimated from the EHT observations. The two red lines in the left plot denote the spin measurement $a/M=0.90\pm 0.05$ from the radio intensity data.}}
    \label{Fig:theta_d}
\end{figure*}

In the five dimensional uniform black string spacetime, the motion of  a neutral test particle with mass $m$ is determined by five geodesic equations characterized by the conserved quantities of the particle: the energy $E_{\rm 0}$, angular momentum $L_{\varphi}$ along the axis of rotation and momentum $P_{\rm z}$ in $z$ direction of the extra dimension. In the first four geodesic equations, the momentum $P_{\rm z}$ always appears in the form of $\sqrt{m^2+P_{\rm z}^2}$, indicating that the momentum along the extra dimension participates as an effective mass $\sqrt{m^2+P_{\rm z}^2}$ for the particle. In particular, for a photon ($m=0$) with a momentum $P_{\rm z}$ along the extra dimension, the motions along the first four dimensions are exactly the same with the motions of a massive particle $m=P_{\rm z}$ without the extra dimension. Also, if a particle can not move along the extra dimension, then we can not recognize the existence of extra dimensions via the motion of particle or shadow.
Therefore,  though many structures of the black string  such as the positions of horizons, ring singularity, ergoregion and causality violation region  are  the same as those for four dimensional  Kerr black hole spacetime, its photon region could be changed due to the existence of extra dimension.  The difference of the photon regions between the five dimensional rotating black string  and the Kerr black hole are  shown in FIG. \ref{Fig:ShadowXY}. With the increase of $P_{\rm E}\equiv P_{\rm z}/E_{\rm 0}$, the photon regions outside the event horizon move out, while the photon regions inside the Cauchy horizon move in and finally the photon regions inside the Cauchy horizon will disappear.

The unstable photon regions outside the event horizon make the direct observation of black holes possible. The photons that escape the spherical photon orbits of a black hole due to the instability and are received by an observer in the domain of outer communication, form the boundary of the dark silhouette of the black hole. This dark silhouette is the so-called black hole shadow from the view of the observer. For each light ray sent from the observer into the past, the initial direction can be described by two angles in the observer's sky, dubbed celestial coordinates. For the black string, an extra angle is produced along the direction of the extra dimension, which can not be perceived by the observers, but the extra dimension indeed have imprints on the celestial coordinates. Using the stereographic projection, it is well known that the shadow boundary in Cartesian coordinates depends on the spin parameter as well as the radial and angle positions of observer. FIG. \ref{Fig:ShadowXY} shows the typical shadow boundary for an observer at spatial infinity in equatorial plane. The motion along the extra dimension apparently enlarges the shadow size or area, while it has slight impact on the shadow distortion. In contrast, the spin parameter mainly distorts the shadow while minishes the shadow area mildly.

In the above study, the length of the extra dimension does not explicitly influence the shadow boundary, which is not anticipated before the calculation, but we could expect to find some clues of the extra dimension using the EHT observations on the supermassive M87* and SgrA* black holes. Especially, the EHT constraints of the shadow angular diameters are $\theta_{\rm d}=42\pm 3 \mu as$ for  supermassive M87* and $\theta_{\rm d}=48.7\pm 7 \mu as$ for  SgrA*, and with the shadow boundary, it is defined as $\theta_{\rm d}=2R_a/d$ with $R_a$ the shadow areal radius and $d$ the distance
of the supermassive black hole from earth. To this end, we could calculate $\theta_{\rm d}$ with the precise measurement of the mass and distance of the black holes and constrain it with the EHT observations.

FIG. \ref{Fig:theta_d} shows the shadow angular diameter of five dimensional black string as the supermassive M87* and the corresponding EHT observations. For fixed $P_{\rm z}/E_{\rm 0}$, the lower bound $\theta_{\rm d}=39~\rm{\mu as}$ could constrain the dimensionless quantity $c P_{\rm z}/E_{\rm 0}=v_{\rm z}/c\lesssim 0.35$ which is the possible maximum value within the spin measurement. However, this constraint could not be proper for an infinite extra dimension because the value of $P_{\rm z}/E_{\rm 0}$ can be arbitrary within $0\sim 1$. In addition, the result shows that the shadow boundary may go beyond the outer border $\theta_{\rm d}=45~\rm{\mu as}$, which indicates that the luminosity distribution produced by these photons is beyond the observed bright region and hence it contradicts with the EHT observations.
We propose that a  {compact extra dimension} can be a suitable option for the particular choices of $P_{\rm z}/E_{\rm 0}$.
The schedule is as follows. For a {compact extra dimension} with length $\ell$, the momentum $P_{\rm z}$ is limited to be box normalized $P_{\rm z} =2\pi\hbar n/\ell~, (~n=0, \pm 1, \pm 2,~...~,)$, such that the length of the extra dimension can be related by the way $v_{\rm z}/c=c P_{\rm z}/E_{\rm 0}=2\pi \hbar n c/(E_{\rm 0} \ell)=n \lambda_{\rm 0}/\ell$ where $\lambda_{\rm 0}$ is the wavelength of the photons. Then, recalling the EHT observing wavelength $\lambda_{\rm 0}= 1.3~mm$, $P_{\rm z}/E_{\rm 0}\lesssim 0.64$ given by upper bound $\theta_{\rm d}=45~\rm{\mu as}$ and the velocity $v_{\rm z}/c \le 1$ only adopt the cases with $n=0$ and $n=0,1$. The former case implies the photons can not move along the extra dimension, and requires the length of the compact extra dimension to be smaller than the wavelength of the photons. While the latter case in turn constrains the length of the extra dimension as $2.03125~\rm{mm} \lesssim \ell \lesssim 2.6~\rm{mm}$, and the parallel analysis from the EHT observations of SgrA* gives a tighter constraint $2.28070~\rm{mm} \lesssim \ell \lesssim 2.6~\rm{mm}$.

In conclusion, we show that the EHT observations of
M87* and SgrA* can not only rule out the black string with an infinite extra
dimension, but also constrain the length of a compact extra dimension, which is
much smaller than the critical length of the static black string to avoid the Gregory-Laflamme
instability \cite{Harmark:2007md}. The finding is surprising for the first consideration since it supports the hypothesis that the extra dimension is compact avoiding the GL instability. The existence and the essence of extra dimension are important open questions in physics and attract plenty of attentions, especially in the content of quantizing the gravity. We hope that our conclusion on extra dimension from the EHT observation can be further verified from other observations, such as the light deflection
experiment in solar system and the gravitational wave.

\section*{Acknowledgments}
We are grateful to Rong-Gen Cai, Yong-Shun Hu and Dong-Chao Zheng for beneficial discussions. This work is supported by National Natural Science Foundation of China (12147119 and 12075202), China Postdoctoral Science Foundation (2021M700142) and Natural Science Foundation of Jiangsu Province (BK20211601).

\end{document}